\documentclass[final,10pt,times,twocolumn]{elsarticle}
\usepackage[margin=1.8cm]{geometry}% by courtesy of Mico
\usepackage{amsmath}
\usepackage{amsfonts}
\usepackage{amssymb}
\usepackage{graphicx}
\usepackage{lmodern}
\usepackage{hyperref}
\usepackage{dcolumn}% Align table columns on decimal point
\usepackage{bm}% bold math
\usepackage[mathlines]{lineno}% Enable numbering of text and display math
\usepackage{float}
\usepackage[T1]{fontenc}
\usepackage{subfigure}
\usepackage{mathtools, cuted}
\usepackage[english]{babel}
\usepackage{lipsum}
\usepackage{siunitx}
\journal{}

\usepackage{numcompress}%\bibliographystyle{model6-num-names}

\DeclareUnicodeCharacter{2212}{\ensuremath{-}}
\DeclareUnicodeCharacter{202F}{FIX ME!!!!}
   \DeclareUnicodeCharacter{2212}{-}

\begin{document}
\newcolumntype{P}[1]{>{\centering\arraybackslash}p{#1}}
\title{\textbf{A Novel Neutrino Mass Matrix}}
%\tnotetext[mytitlenote]{Fully documented %templates are available in the elsarticle %package on \href{http://www.ctan.org/tex-%archive/macros/latex/contrib/elsarticle}{CTAN}.}



\author[mymainaddress]{Pralay Chakraborty}
\ead{pralay@gauhati.ac.in}

\author[mymainaddress]{Sagar Tirtha Goswami}
\ead{sagartirtha@gauhati.ac.in}

\author[mymainaddress]{Subhankar Roy\corref{mycorrespondingauthor}}
\cortext[mycorrespondingauthor]{Corresponding author}
\ead{subhankar@gauhati.ac.in}

\address[mymainaddress]{Department of Physics, Gauhati University, India}


\begin{abstract}
A predictive neutrino mass matrix texture, sheltering unique correlations ($m_{12}=m_{13}\,\,\&\,\,m_{33}=2i\, m_{12}$), is proposed, addressing most of the timely neutrino phenomenology issues. The texture is realized in the framework of type-I + type-II seesaw in the light of the $A_4 \times Z_{10} \times Z_{7} \times Z_{3}$ group. The stability of the proposed texture is studied under renormalization group evolution.
\end{abstract}
\maketitle

The quest to resolve the open problems in neutrino physics through the study of mass matrix textures, using both bottom-up and top-down approaches, remains highly relevant. Constructing a predictive neutrino mass matrix with the minimum number of parameters and deriving it from a robust theoretical framework is undoubtedly a challenging task. Although extensive work exists in the literature\,\cite{Altarelli:2005yx,Ludl:2014axa,Kalita:2015tda,Chakraborty:2024eki,Chakraborty:2022ess,Ismael:2021jay,Goswami:2023eyy}, the search for viable textures continues, since experiments have not fully unlocked the mysteries of the physical parameters. In sum, neutrino phenomenology is depicted in terms of nine physical parameters: namely, three neutrino mass eigenvalues $m_{i=1,2,3}$; three mixing angles, $\theta_{13}$, $\theta_{12}$, and $\theta_{23}$, signifying the mixing among neutrino flavour and mass eigenstates; and the Charge-Parity (CP) violating phases $\delta$, $\alpha$, and $\beta$. These CP phases are important in the sense that they carry the signature of whether neutrinos are Dirac or Majorana particles. So far, the oscillation experiments are sensitive only to $\delta$. The parameters $\alpha$ and $\beta$ are exclusively related to the Majorana nature of neutrinos. However, the present work presumes the neutrino to be a Majorana particle. Neutrino oscillation experiments measure the two neutrino mass-squared differences\,($\Delta m^2_{21}$ and $\Delta m^2_{31}$) and not the individual $m_i$'s. Moreover, the sign of $\Delta m^2_{31}$ is unknown, leaving open the possibilities of Normal Hierarchy ($m_3 > m_2 > m_1$)\,(NH) and Inverted Hierarchy ($m_2 > m_1 > m_3$)\,(IH). Within these scenarios, the limiting cases such as $m_1 = 0$ (NH) or $m_3 = 0$ (IH) are also viable. The three mixing angles ($\theta_{12}$, $\theta_{13}$, $\theta_{23}$), though witnessed by the oscillation experiments\,\cite{Esteban:2024eli, ParticleDataGroup:2022pth}, still leave the octant of $\theta_{23}$ as an open question. In addition, a wide range of $\delta$ is consistent with experiments. Because of their inherent nature, oscillation experiments do not address the remaining two CP phases $\alpha$ and $\beta$, whereas other experiments are trying to detect their indirect signatures\,\cite{ KamLAND-Zen:2016pfg, LEGEND:2021bnm}. Although, the existing mysteries in the neutrino sector make the model building a exhaustive task, yet prevents it to reach the saturation as the experiments are continuously reaching precision. The information on all these observable parameters is embedded within a mathematical structure called the neutrino mass matrix. The predictions of the observables are manifested as correlations among the matrix elements. This motivation leads us to posit an efficient neutrino mass matrix bearing four independent complex parameters, equally relevant in terms of present oscillation experiments and its testability in future experiments, as in the following,

\begin{figure*}
  \centering
    \subfigure[]{\includegraphics[width=0.32\textwidth]{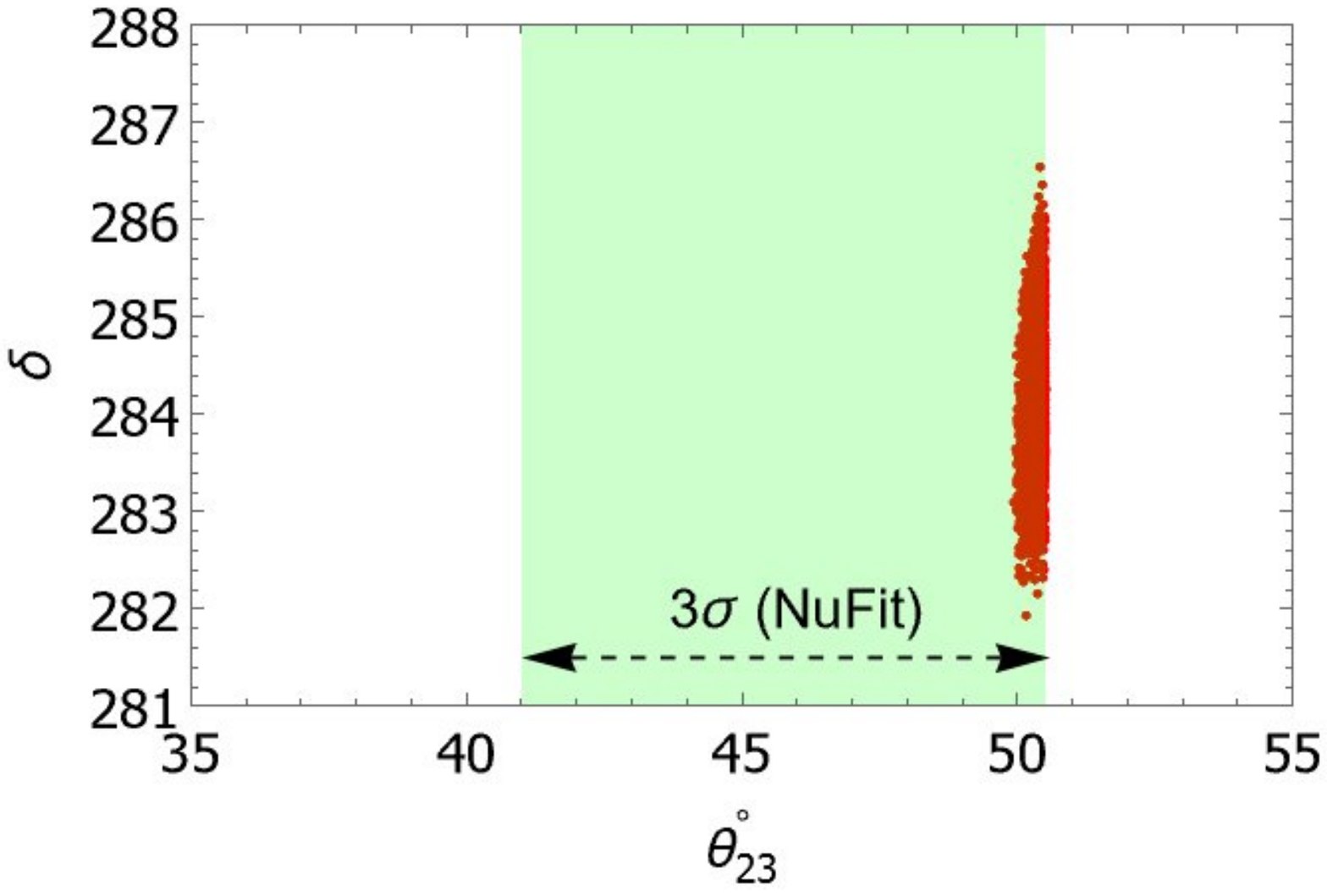}\label{fig:1a}} 
    \subfigure[]{\includegraphics[width=0.32\textwidth]{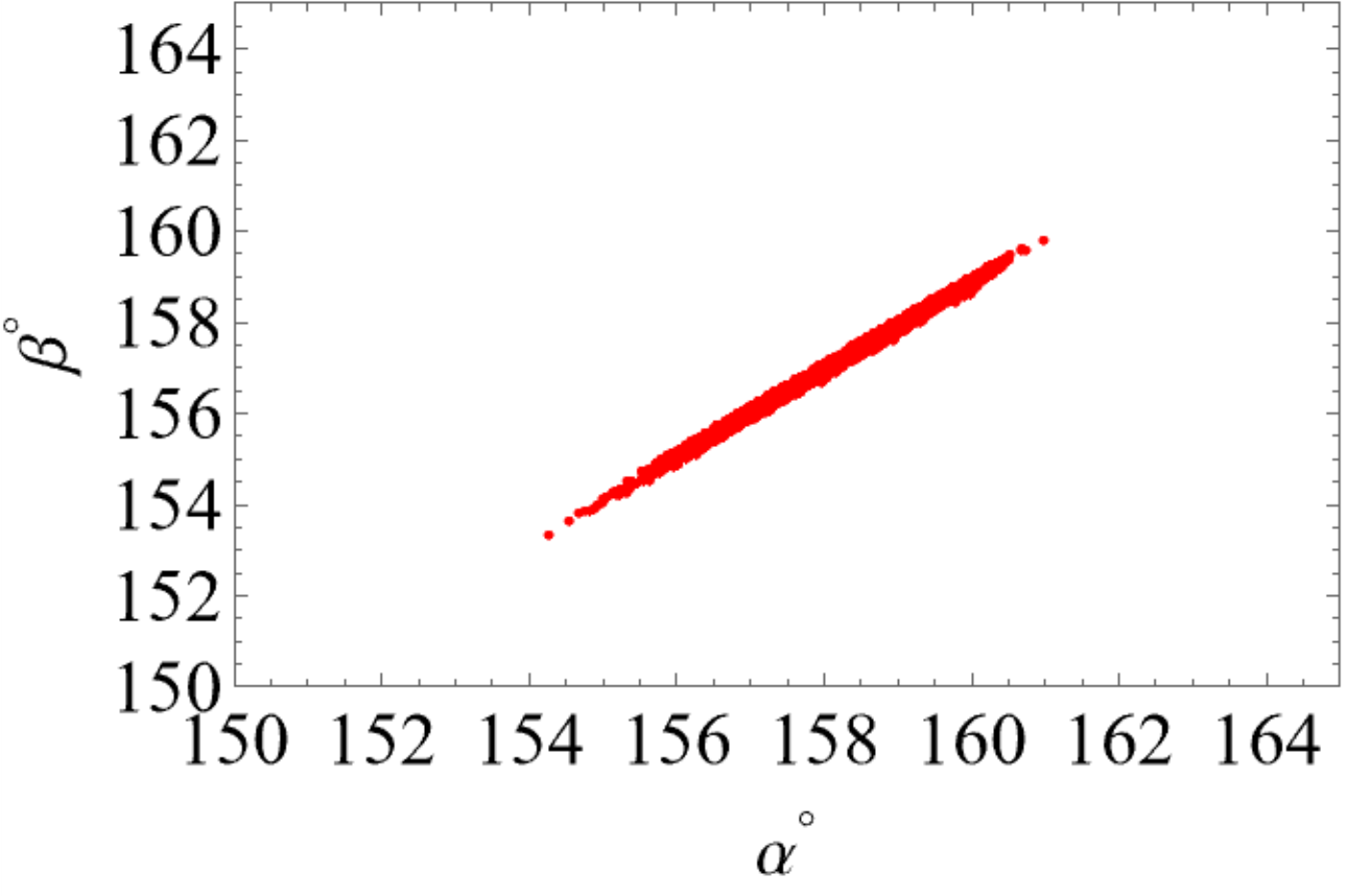}\label{fig:1b}}
    \subfigure[]{\includegraphics[width=0.32\textwidth]{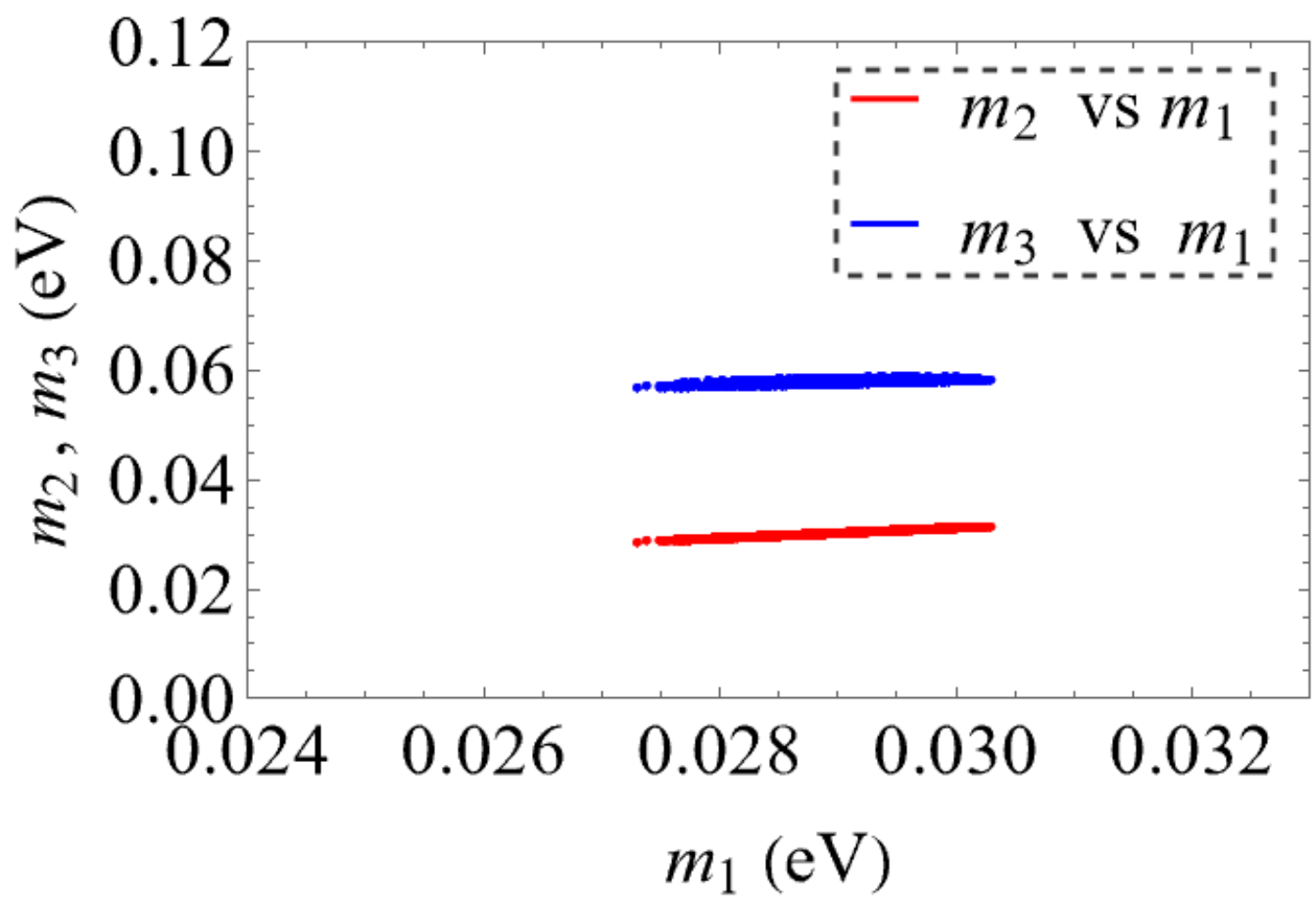}\label{fig:1c}} 
    \subfigure[]{\includegraphics[width=0.32\textwidth]{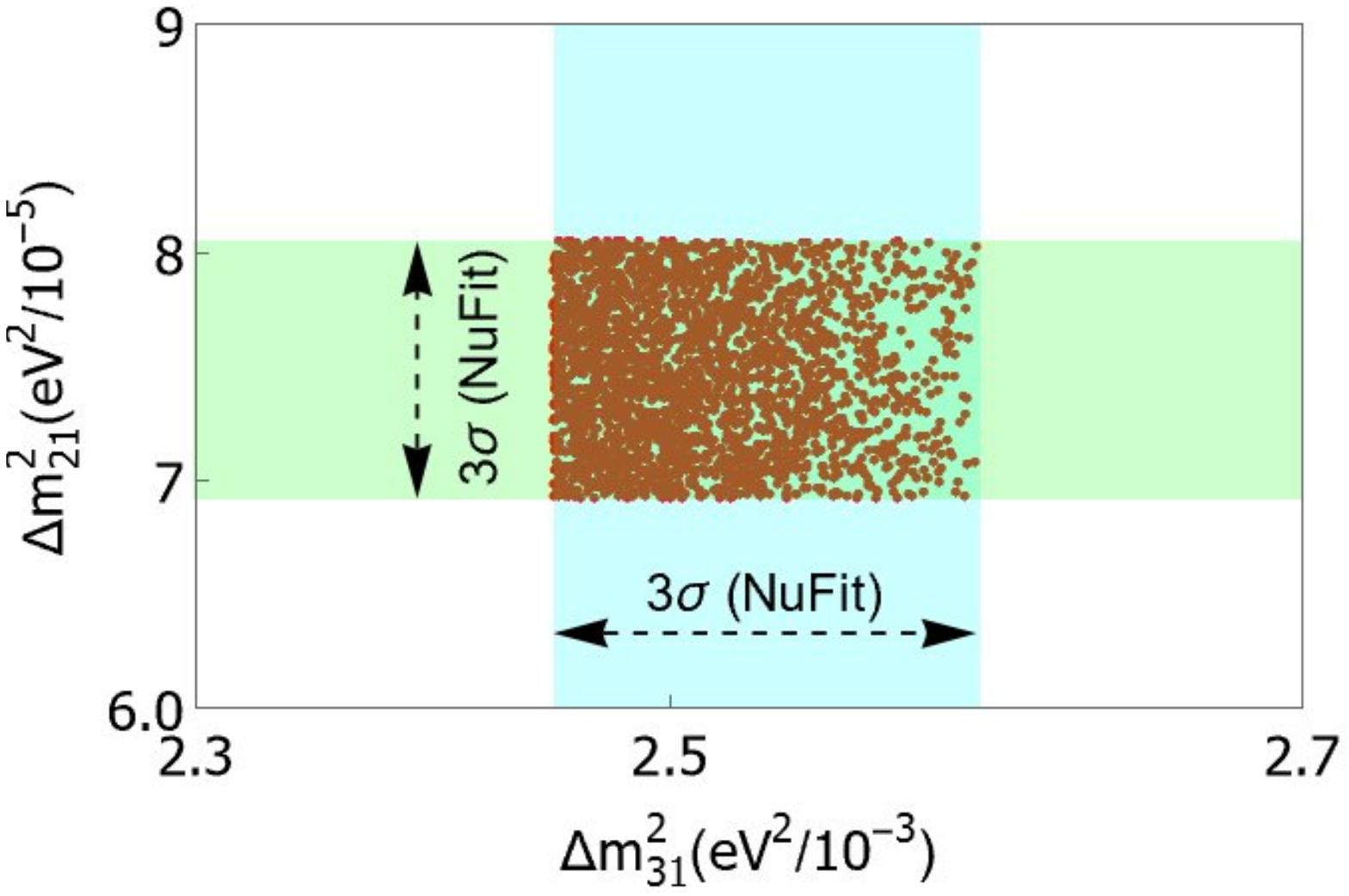}\label{fig:1d}}
    \subfigure[]{\includegraphics[width=0.32\textwidth]{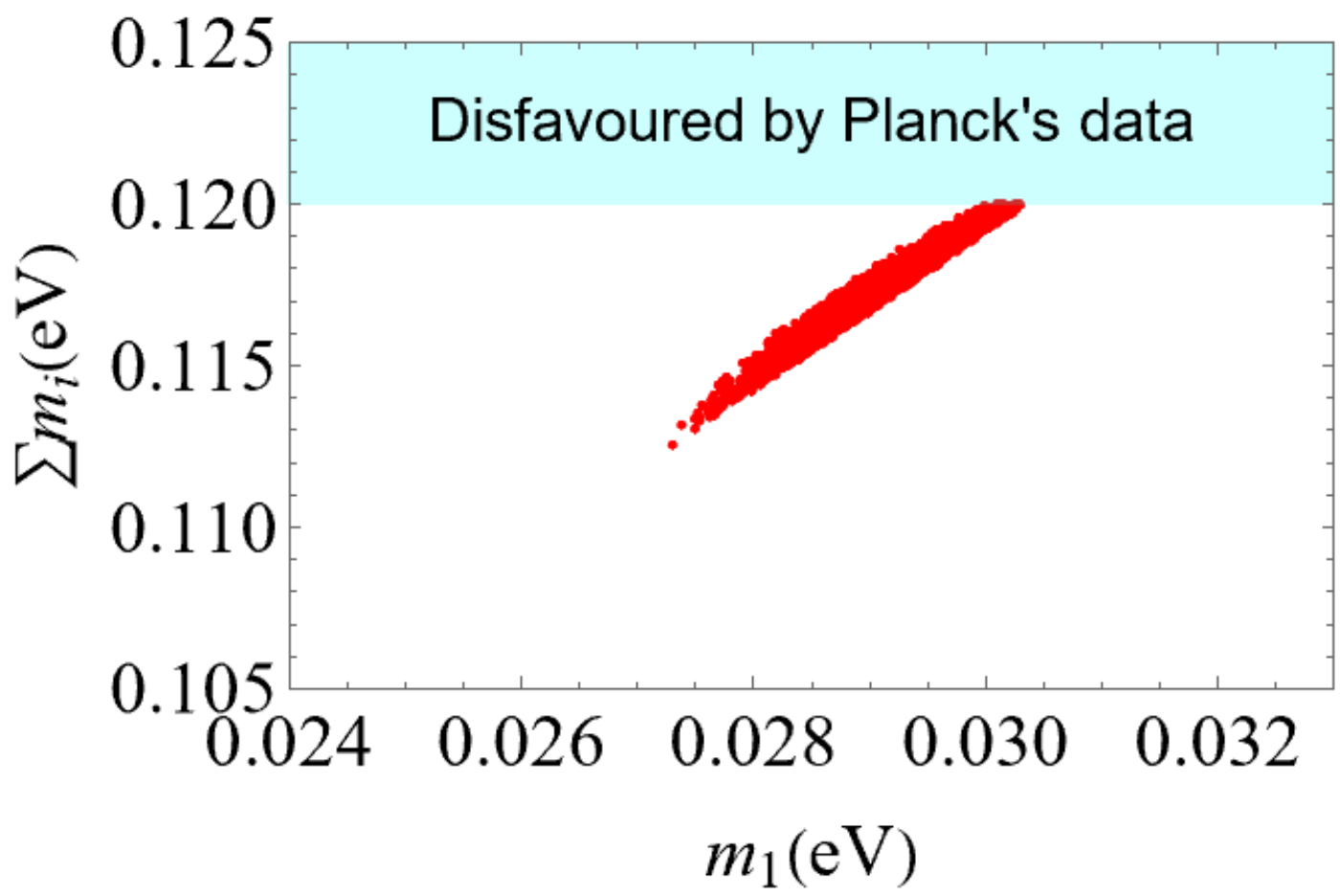}\label{fig:1e}} 
    \subfigure[]{\includegraphics[width=0.34\textwidth]{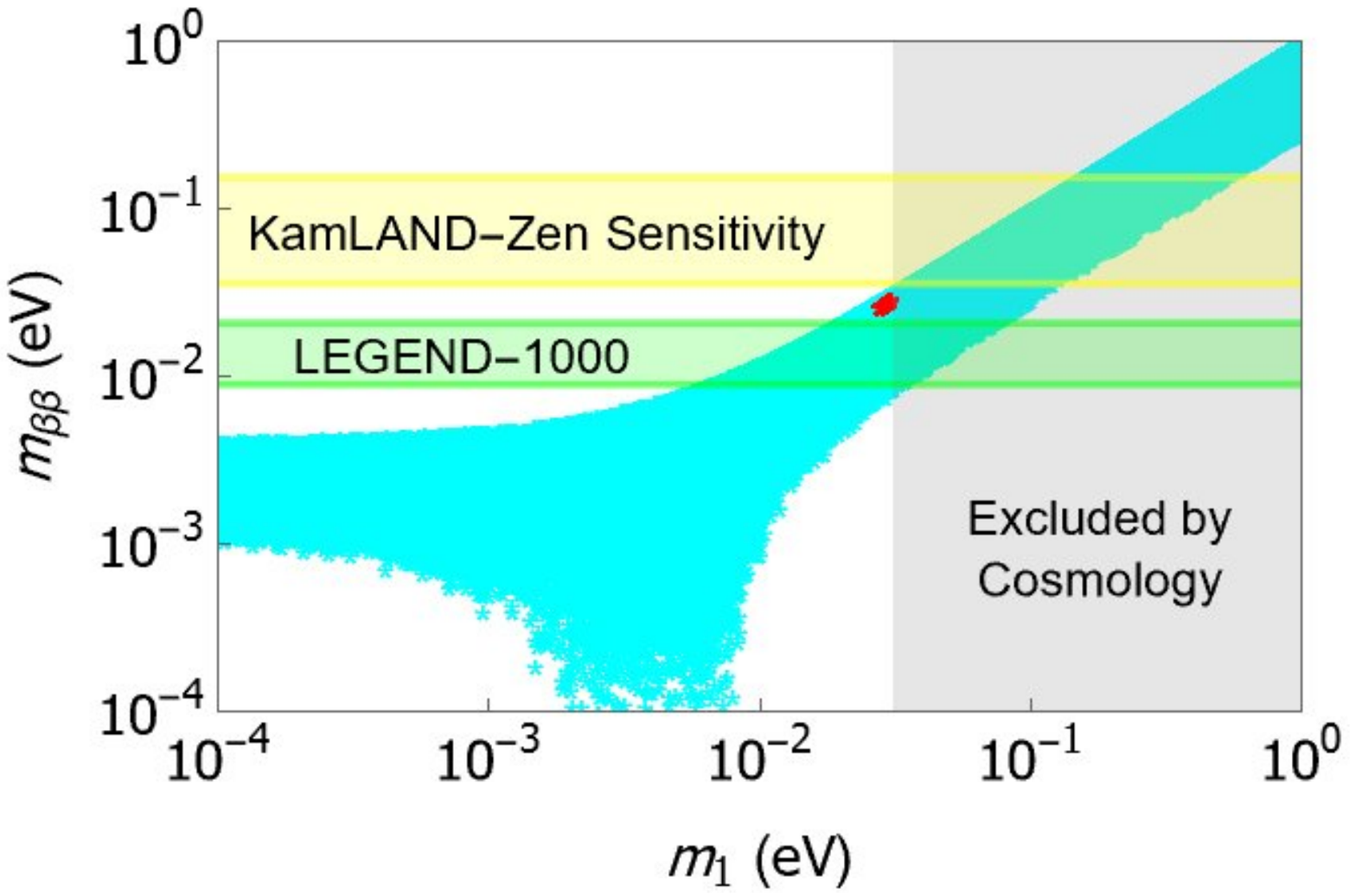}\label{fig:1f}} 
   \caption{ The correlation plots between (a) $\theta_{23}$ vs $\delta$, (b) $\alpha$ vs $\beta$, (c) $m_1$ and $m_2$ vs $m_3$, (d) $\Delta\,m_{21}^2$ vs $\Delta\,m_{31}^2$, (e) $\sum m_i$ vs $m_1$, (f) The effective Majorana neutrino mass $m_{\beta\beta}$ vs $m_1$. }
\label{fig:mass parameters}
\end{figure*}

\begin{figure*}
  \centering
    \subfigure[]{\includegraphics[width=0.32\textwidth]{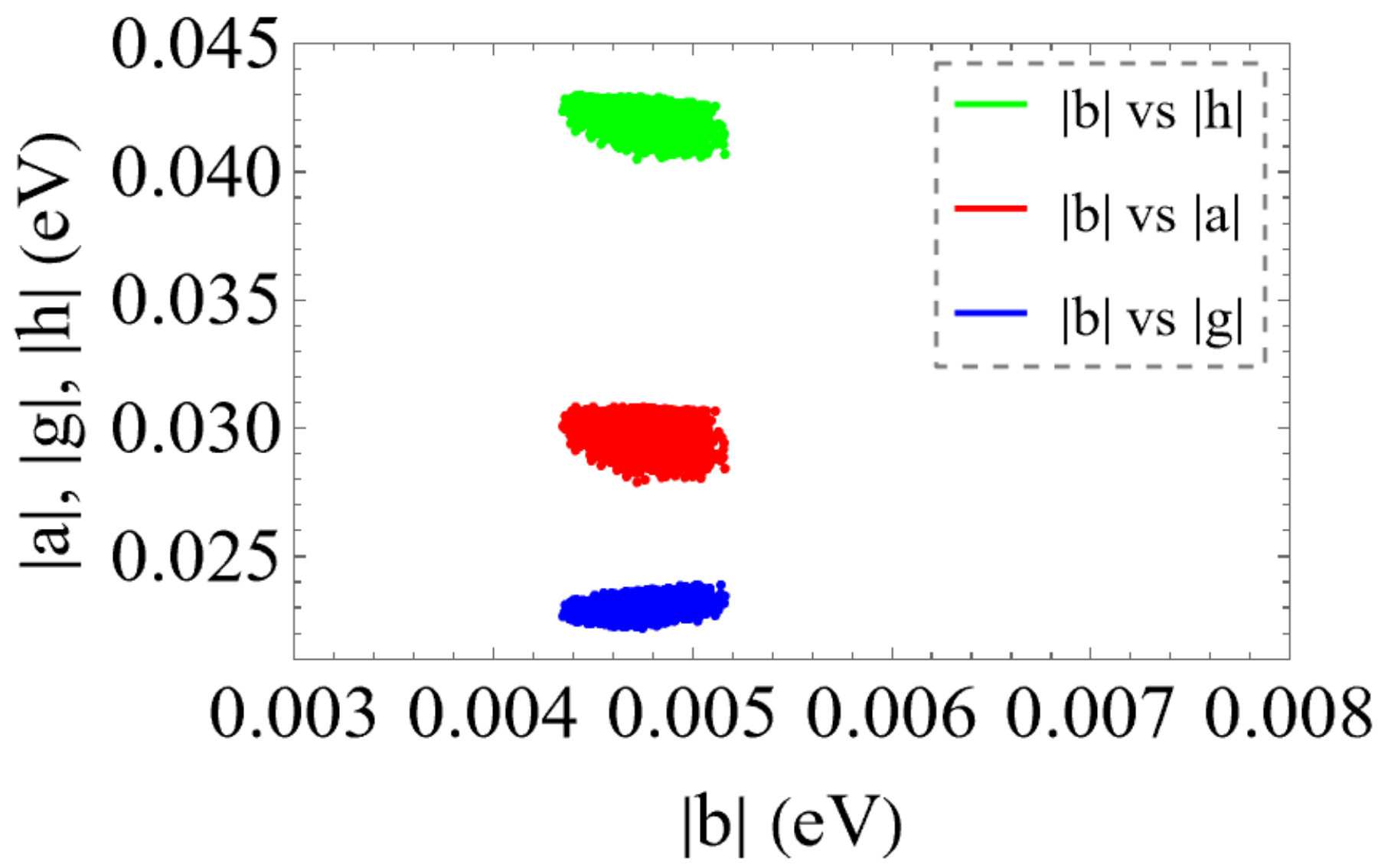}\label{fig:3a}}
    \subfigure[]{\includegraphics[width=0.32\textwidth]{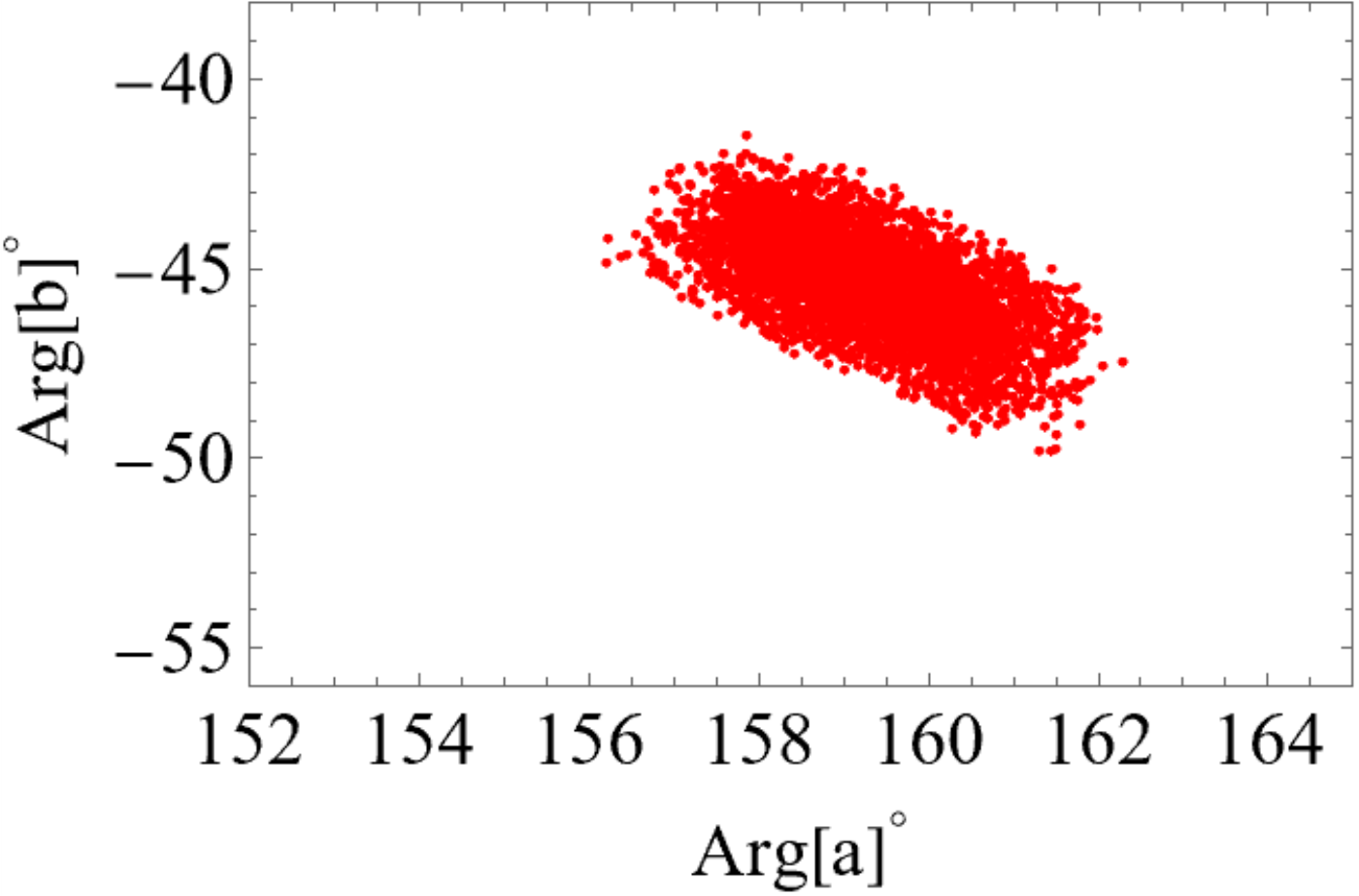}\label{fig:3b}} 
    \subfigure[]{\includegraphics[width=0.32\textwidth]{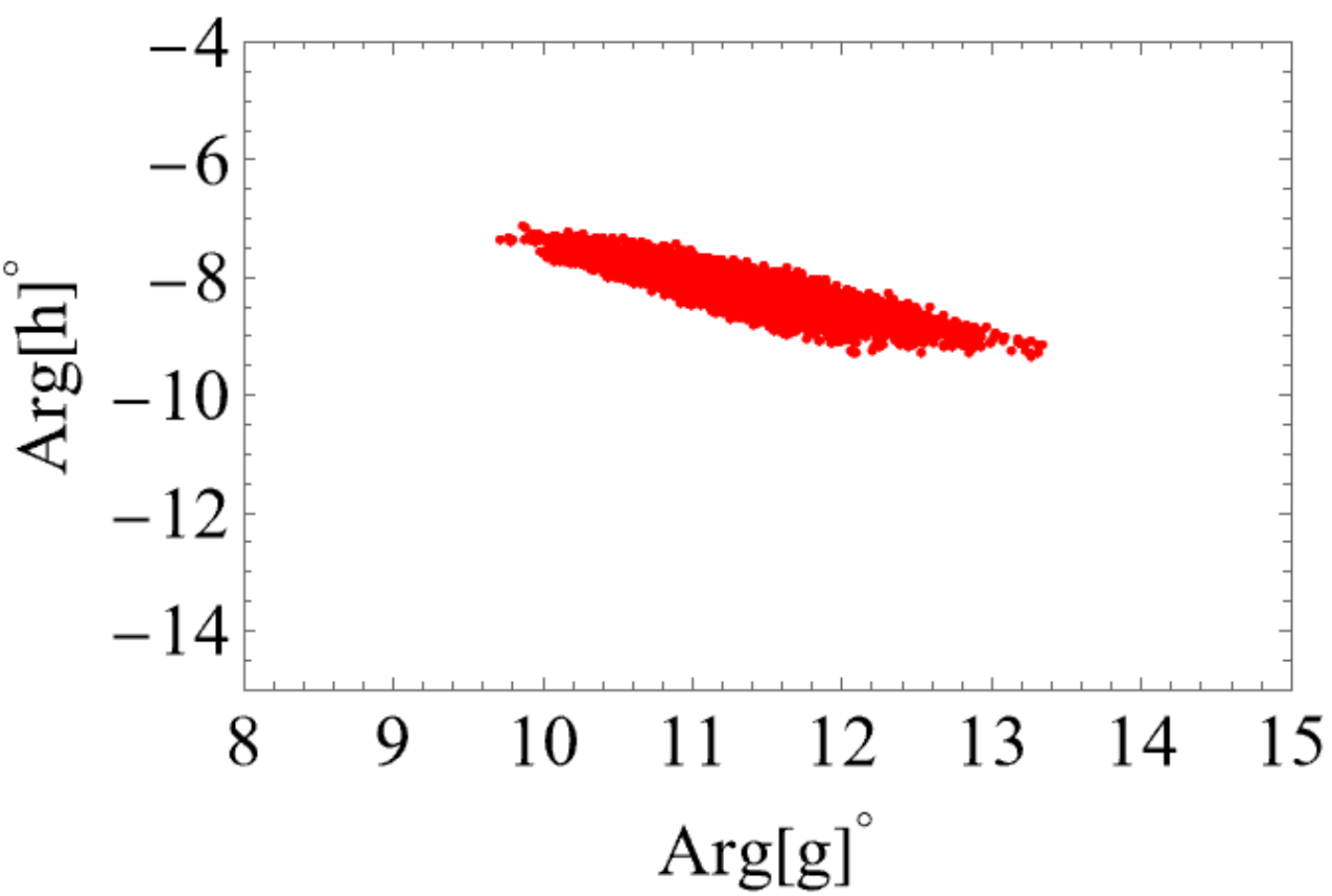}\label{fig:3c}} 
   \caption{ The correlation plots between (a) $|a|$, $|g|$, $|h|$ vs $|b|$, (b) $\text{Arg}[b]$ vs $\text{Arg}[a]$, (c) $\text{Arg}[h]$ vs $\text{Arg}[g]$. }
\label{fig:texture parameters}
\end{figure*}

\begin{figure*}
  \centering
    \subfigure[]{\includegraphics[width=0.32\textwidth]{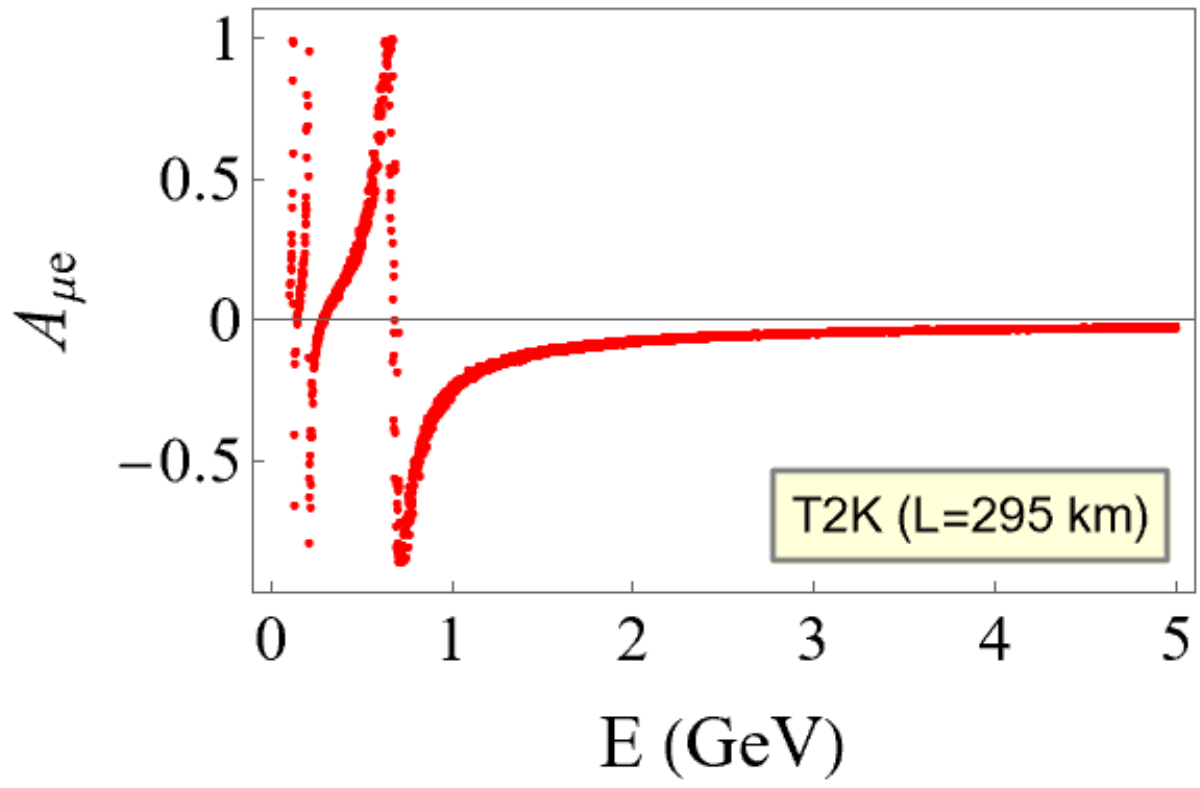}\label{fig:2(a)}} 
    \subfigure[]{\includegraphics[width=0.32\textwidth]{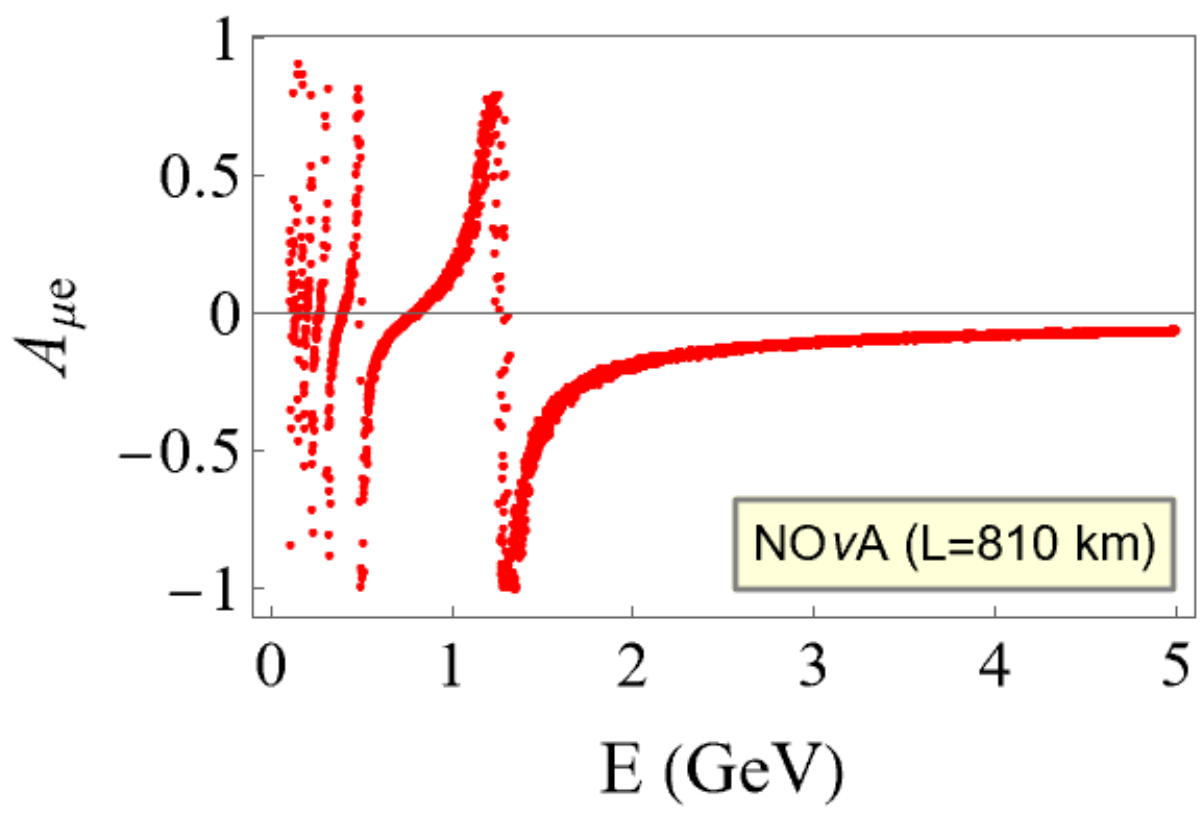}\label{fig:2(b)}} 
    \subfigure[]{\includegraphics[width=0.32\textwidth]{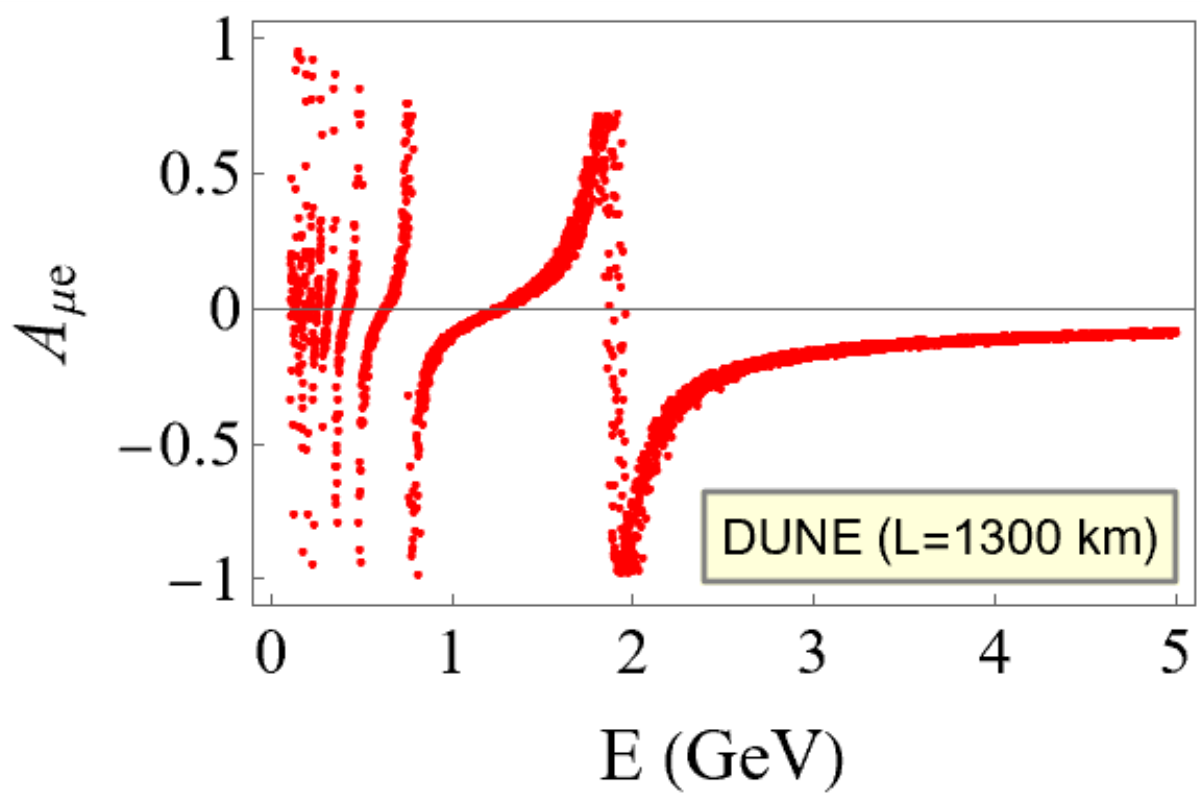}\label{fig:2(c)}}
    \caption{The correlation plots between (a) $A_{\mu e}$ vs $E$ for T2K $(L=295)$ km, (b) $A_{\mu e}$ vs $E$ for NO$\nu$A $(L=810)$ km, (c) $A_{\mu e}$ vs $E$ for DUNE $(L=1300)$ km.}
\label{fig:1}
\end{figure*}

\begin{equation}
\label{Mnu}
M_\nu = \begin{bmatrix}
a & b & b \\
 b & g & h\\
 b & h & 2ib
\end{bmatrix},
\end{equation}

highlighting two unique correlations that have not been explored earlier in the literature. The proposed texture exhibits several interesting features. For example, it constrains $\theta_{23}$ and three CP phases within stringent bounds. In addition, the texture rules out the inverted hierarchy of neutrino masses predicting three neutrino mass eigenvalues. At first glance, the mass matrix shown in Eq.\,(\ref{Mnu}) may appear arbitrary and it is important to find it's novel theoretical origin. Interestingly, we observe that the proposed texture can be realized in its exact form starting from a discrete symmetry based seesaw framework.

\begin{table}[b]
\centering
\begin{tabular*}{0.47\textwidth}{@{\extracolsep{\fill}} ccc}
\hline
\hline
Parameters & Minimum Value &  Maximum Value \\
\hline
\hline
$\theta_{23} /^\circ$ & 49.90 & 50.02 \\
\hline
$\delta /^\circ$ & 281.93 & 286.53 \\
\hline
$\alpha/^\circ$ & 154.26  & 160.97  \\
\hline
$\beta/^\circ$ & 153.31 & 159.80  \\
\hline
$m_1$/eV & 0.027 & 0.030 \\
\hline
$m_2$/eV & 0.028 & 0.031  \\
\hline
$m_3$/eV & 0.056 & 0.058  \\
\hline
\end{tabular*}
\caption{Shows the maximum and minimum values of  $\theta_{23}$, $\delta$, $\alpha$, $\beta$, $m_1$, $m_2$ and $m_3$ predicted from the proposed texture.}
\label{physical parameters}
\end{table}

\begin{table*}[t]
\centering
\begin{tabular*}{\textwidth}{@{\extracolsep{\fill}} cccccccccccc}
\hline
\hline
Fields & $D_{l_{L}}$ & $l_{R}$ & $\nu_{l_R}$ & $H$ & $\Delta$ & $\chi$ & $\psi$ & $\xi$ & $\eta$ &  $\kappa$ & $\rho$ \\ 
\hline\hline
$SU(2)_{L}$ & 2 & 1 & 1 & 2 & 3 & 1 & 1 & 1 & 1 & 1 & 1 \\
\hline
$U(1)_Y$ & -1 & $(-2,-2,-2)$ & $(0,0,0)$ & $1$ & -2& 0 & 0 & 0 & 0 & 0  & 0 \\
\hline
$A_4$ & 3 & $(1,1',1'')$ & $(1,1',1^{''})$ & 3 & 3 & 1 & 1 & 1 & $1'$ & 1 & 1 \\
\hline
$Z_{10}$ & 0 & $(1,4,7)$ & $(0,4,3)$ & $0$ & 0 & 6 & $7$ & 1 & $2$ & 3 &0 \\
\hline
$Z_{7}$ & -2 & $(4,4,4)$ & $(6,6,6)$ & $1$ & 3 & 0 & $0$ & 0 & $2$ & 0 & 2 \\
\hline
$Z_{3}$ & $\omega^2$ & $(\omega,\omega,\omega)$ & $(1,1,1)$ & $\omega$ & $\omega$ & 1 & $1$ & $\omega$ & $1$ & 1 &1 \\
\hline
\end{tabular*}
\caption{The transformation properties of various fields under $SU(2)_L \times U(1)_Y \times A_4 \times Z_{10} \times Z_{7} \times Z_{3}$ group.} 
\label{Field Content of M}
\end{table*} 

 We extract the information of the observable parameters from $M_\nu$\,\cite{Chakraborty:2024eki, Goswami:2023eyy}. Interestingly, the texture resolves the octant degeneracy of $\theta_{23}$, predicting it to be around $50^{\circ}$. Similarly, the CP phase $\delta$ is also predicted precisely within the fourth quadrant. On the other hand, $\alpha$ and $\beta$ are placed within the second quadrant. The proposed mass matrix rules out any possibility of an inverted hierarchy occurring, further restricting the extreme case of $m_1$ being zero. We highlight that $\Delta\,m_{21}^2$ and $\Delta\,m_{31}^2$ are consistent with the experimental $3\sigma$ bound\,\cite{Esteban:2024eli}, and $\sum m_i$ lies below the observed upper bound $0.12$ eV obtained from cosmological data\,\cite{Planck:2018vyg}. The graphical analysis can be found in Figs.\,((\ref{fig:1a})-(\ref{fig:1e})). The minimum and maximum values of the studied parameters are listed in Table\,(\ref{physical parameters}). A graphical visualization of the said texture parameters is highlighted in Fig.\,(\ref{fig:texture parameters}).

As mentioned earlier, the $\alpha$ and $\beta$ are not direct observables. However, their presence influence intensely the effective Majorana neutrino mass\,\( m_{\beta\beta} \). So, as n potential outcome of Eq\,(\ref{Mnu}), we evaluate it as, $(25.7-28.97)$ meV (see Fig.\,(\ref{fig:1f})) that lies between the sensitivities of KamLAND-Zen and LEGEND-1000 \cite{KamLAND-Zen:2016pfg, LEGEND:2021bnm}. This extends a scope to test the proposed texture in more sophisticated future experiments. In addition, the stringent bound on $\delta$ imposed by the proposed texture motivates us to visualize the variation of the CP asymmetry parameter $A_{\mu e}$\,\cite{Sinha:2018xof}. It is done for three long-baseline neutrino experiments with different baseline lengths: T$2$K\,($L=295$ km)\,\cite{T2K:2019bcf}, No$\nu$A\,($L=810$ km)\,\cite{NOvA:2021nfi} and DUNE\,($L=1300$ km)\,\cite{DUNE:2020ypp}\, (see Figs.\,(\ref{fig:2(a)})-(\ref{fig:2(c)})). For this study, the neutrino beam energy \( E \) is varied within the range \( 0.5~\text{GeV} \leq E \leq 5~\text{GeV}\). Needless to mention, the long-baseline experiments aim to measure this phase while accounting for sub-dominant matter effects. Notably, the results from T2K and NO\(\nu\)A  show differing predictions for \(\delta\) for the inverted ordering\,\cite{T2K:2019bcf,NOvA:2021nfi}. Future experiment such as DUNE with \(L = 1300\)\,km, and a peak beam energy of approximately 2.5 GeV\,\cite{DUNE:2020ypp}, is expected to shed light on this discrepancy, providing a potential test for our prediction of \(A_{\mu e}\).

\begin{figure*}
  \centering
    \subfigure[]{\includegraphics[width=0.32\textwidth]{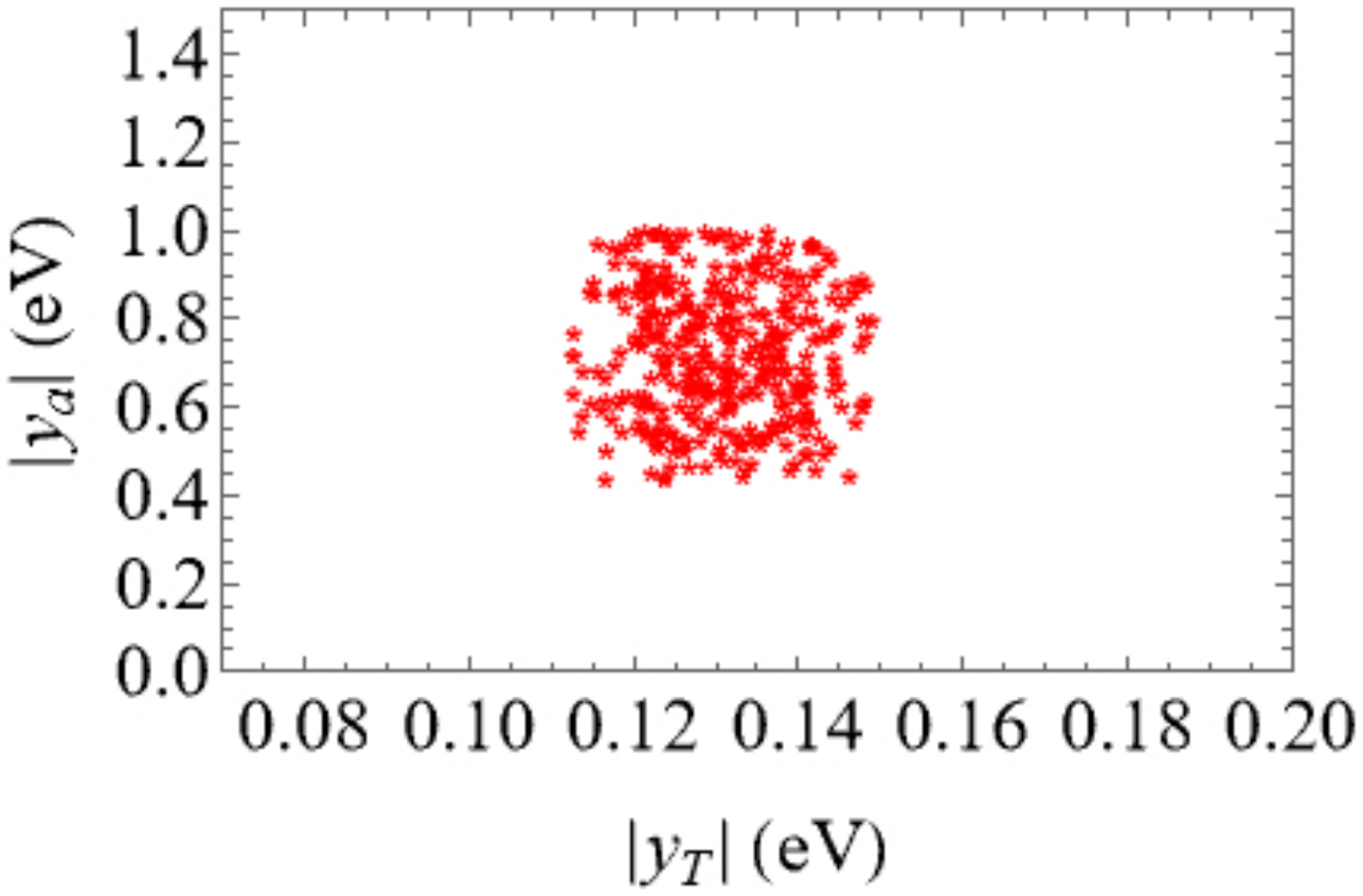}} 
    \subfigure[]{\includegraphics[width=0.32\textwidth]{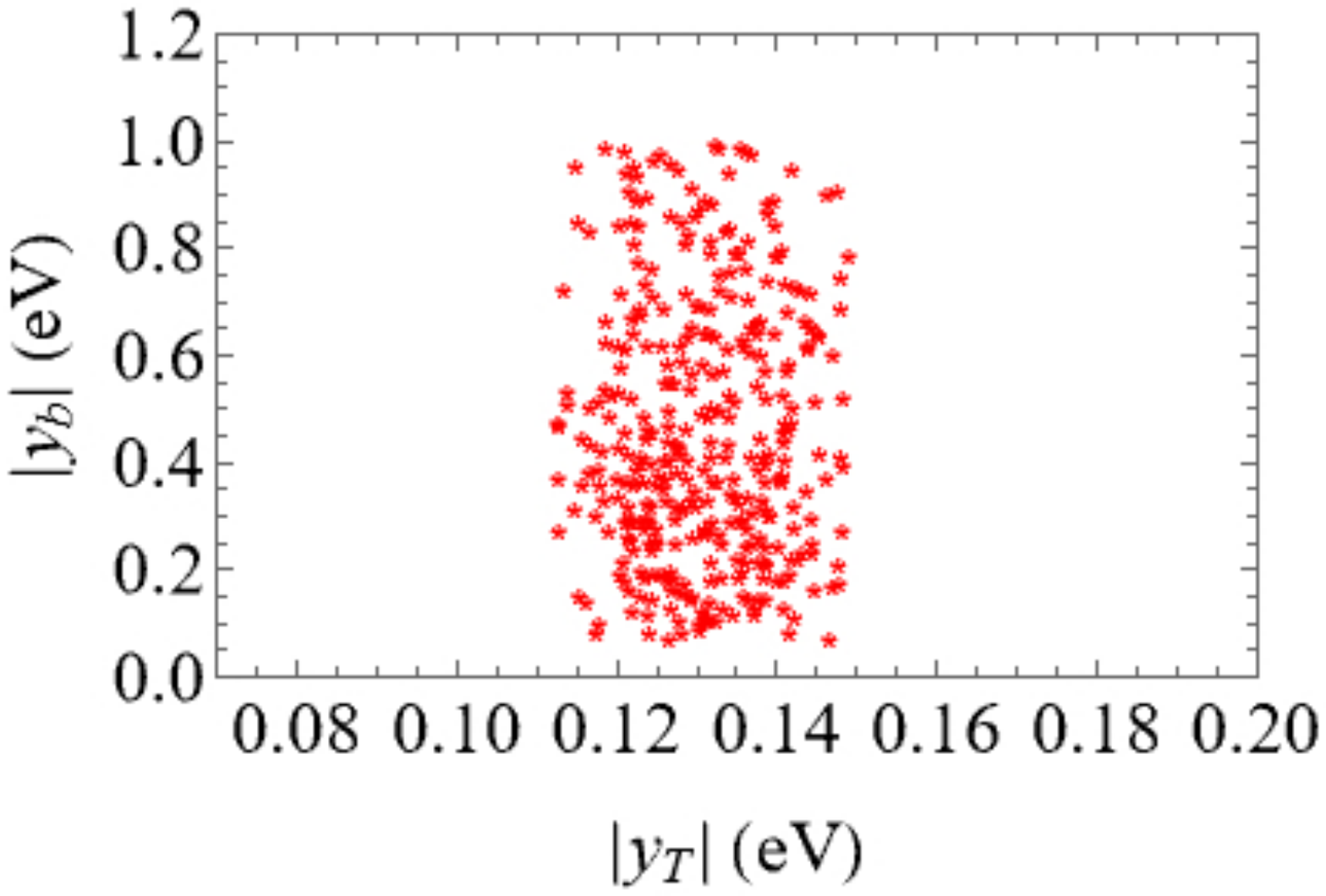}}
    \subfigure[]{\includegraphics[width=0.32\textwidth]{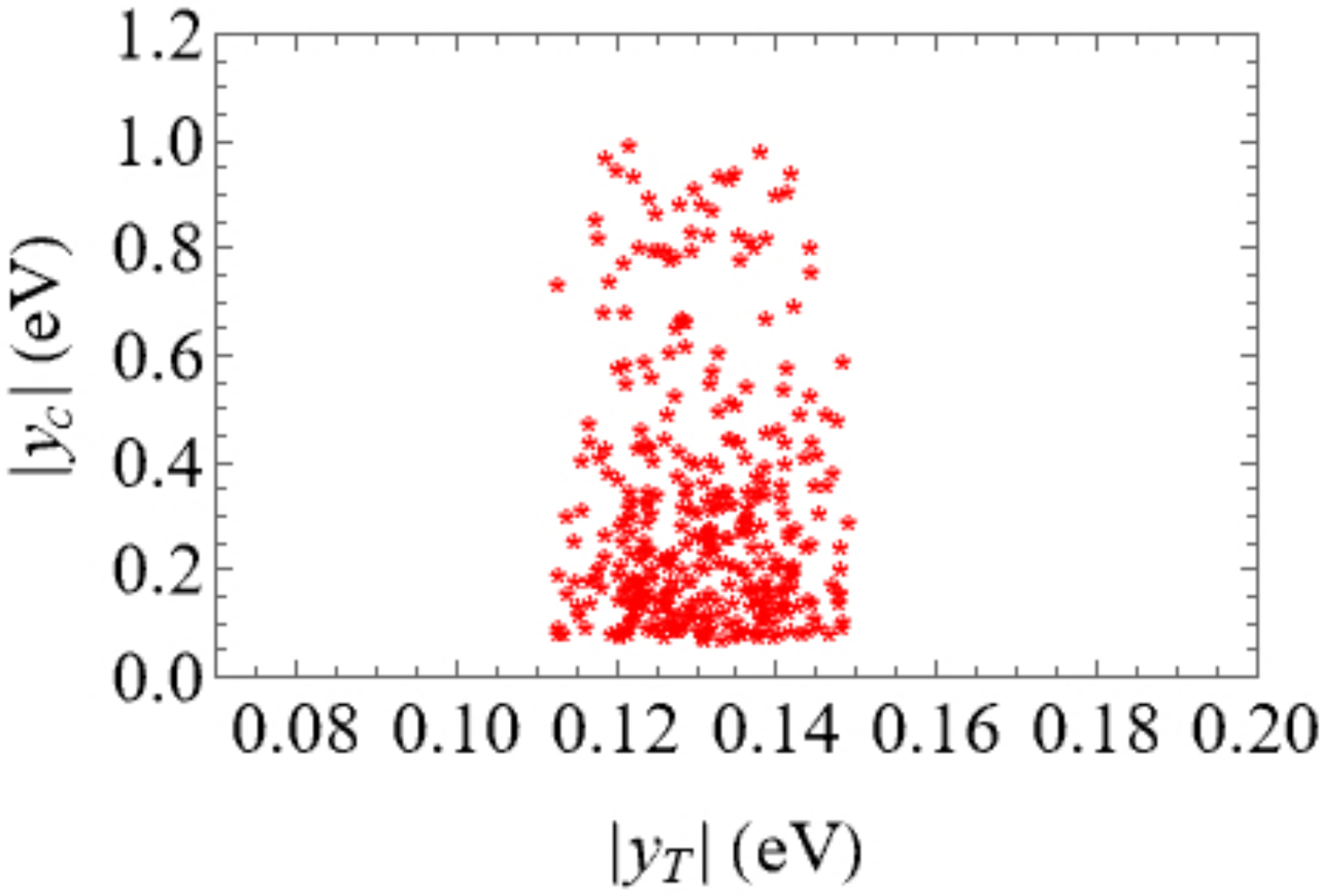}} 
    \subfigure[]{\includegraphics[width=0.32\textwidth]{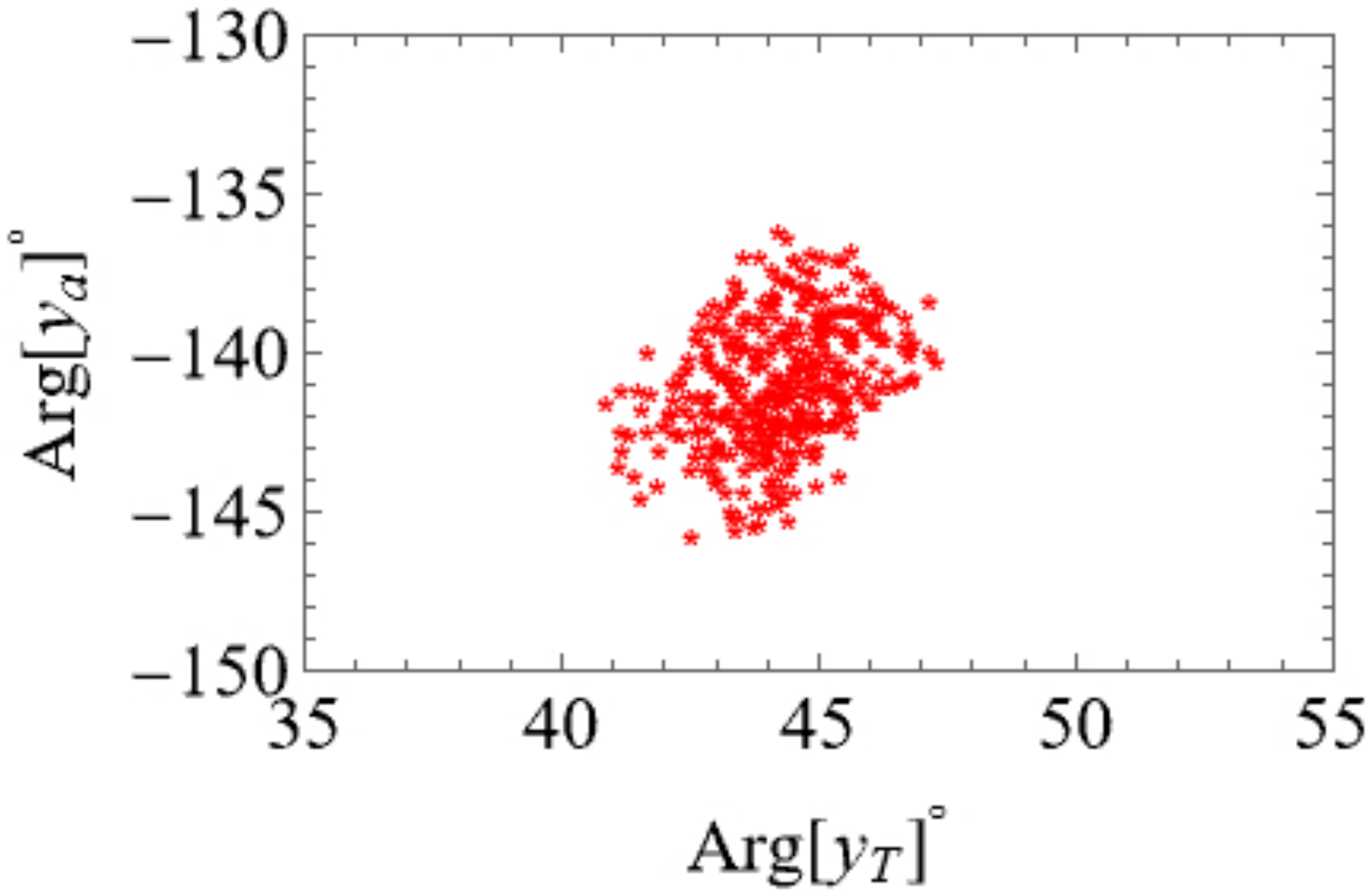}}
    \subfigure[]{\includegraphics[width=0.32\textwidth]{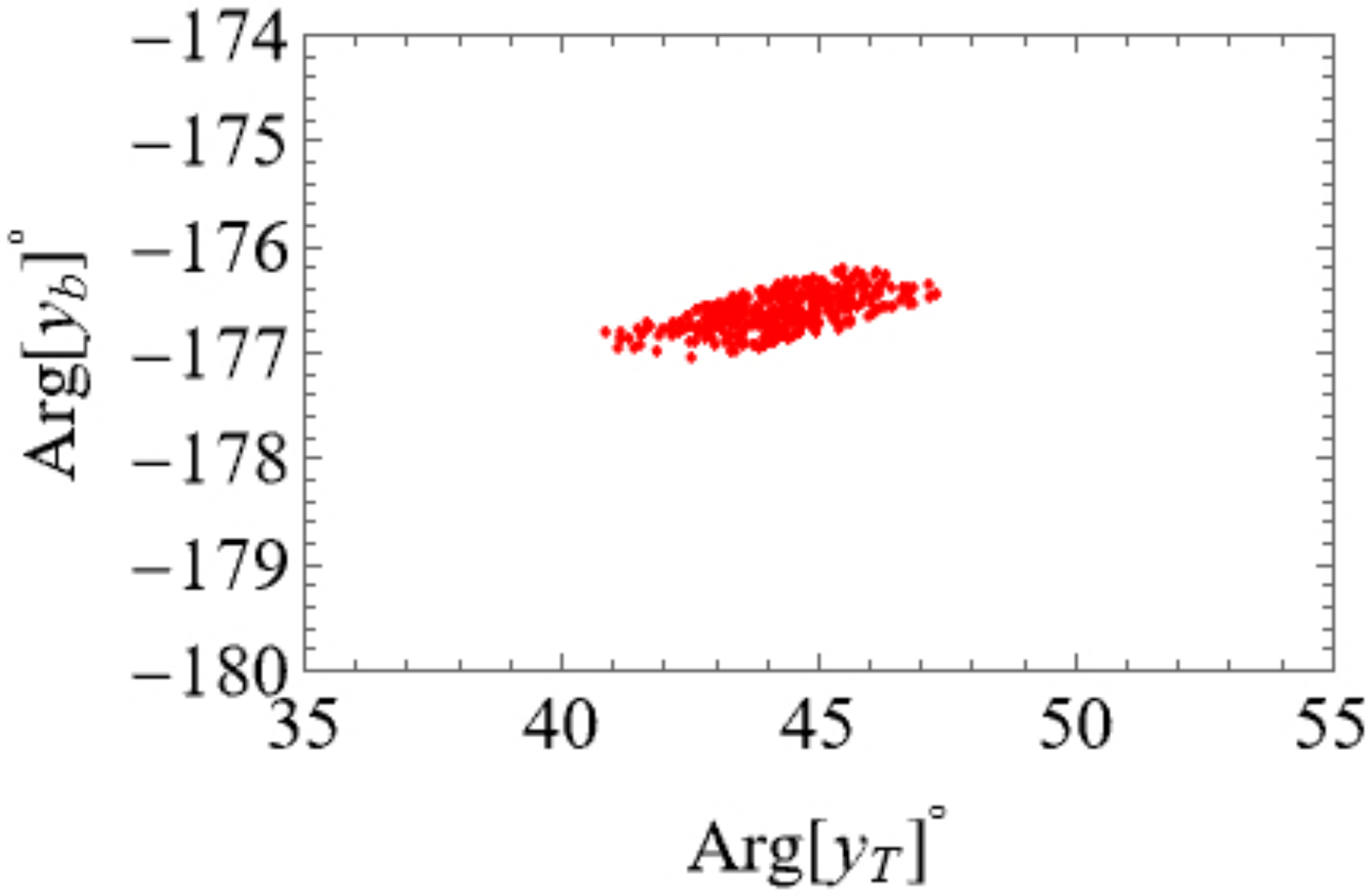}} 
    \subfigure[]{\includegraphics[width=0.34\textwidth]{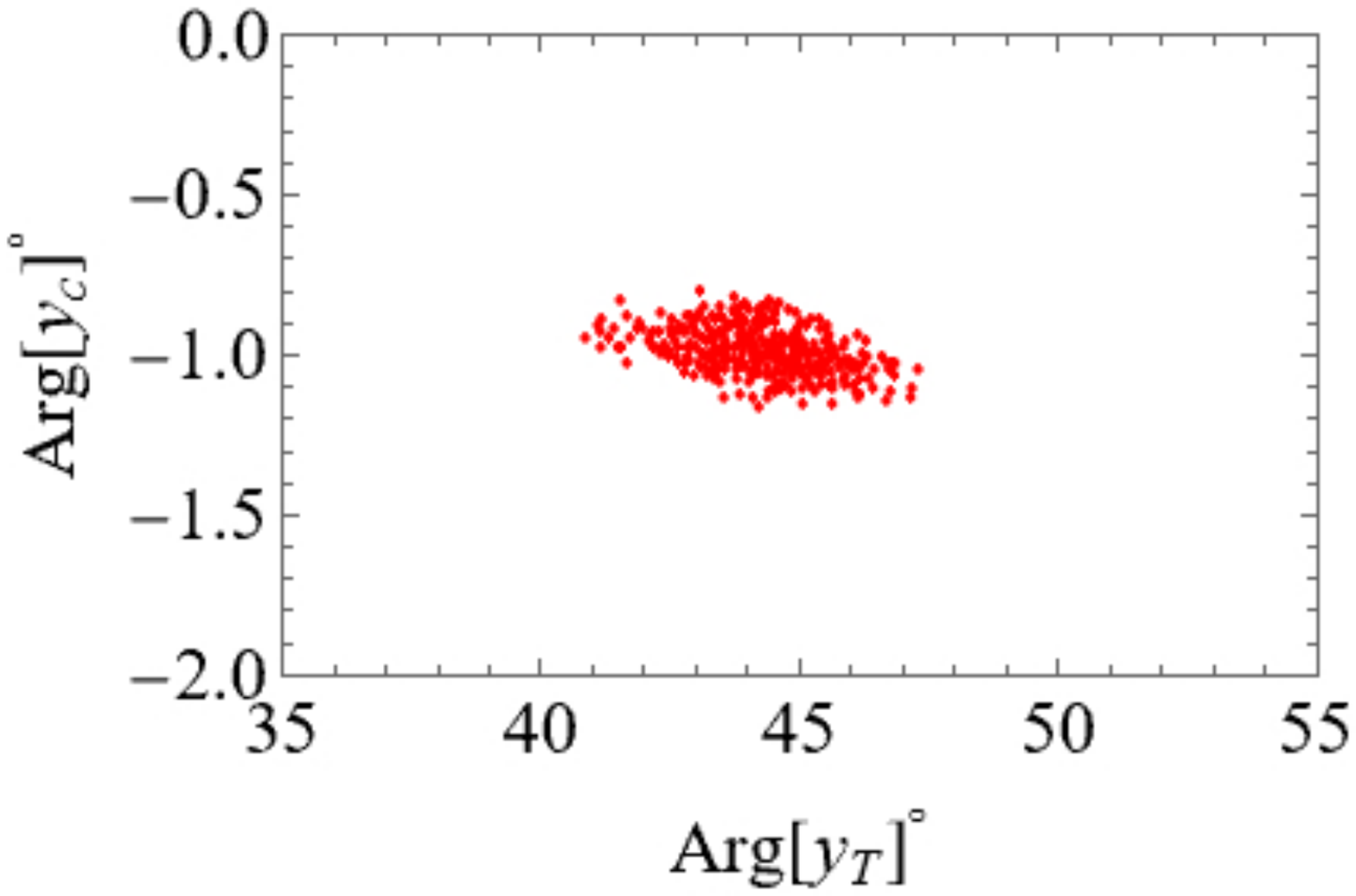}} 
   \caption{ The possible parameter space for the Yukawa couplings. }
\label{fig:yukawa}
\end{figure*}

\begin{figure}
\centering
\includegraphics[scale=0.3]{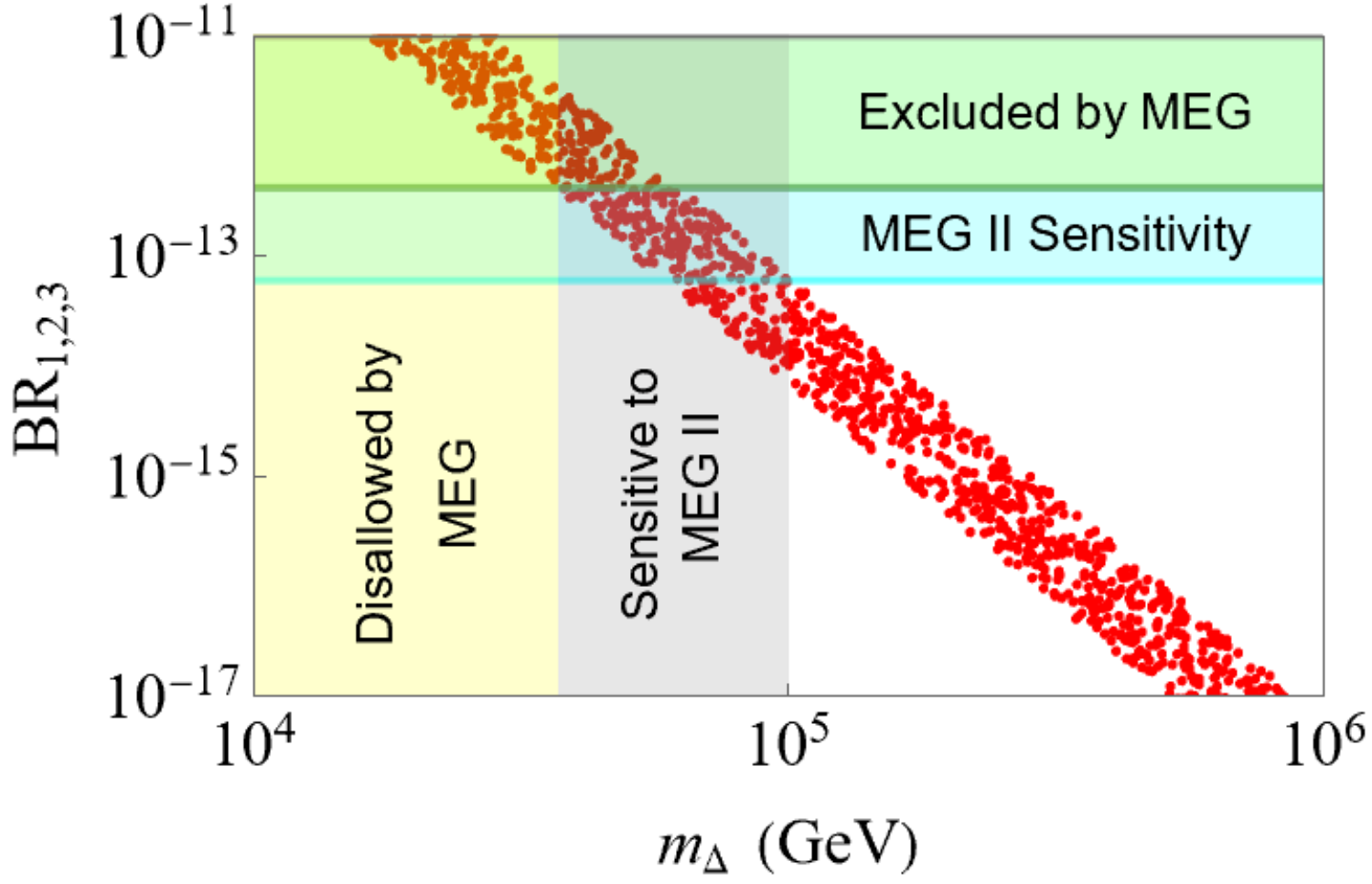}
 \caption{ The variation of $\text{BR}_{1}$ $(\mu \longrightarrow e \gamma)$,\,\,$\text{BR}_{2}$ $(\tau \longrightarrow e \gamma)$ and $\text{BR}_{3}$ $(\tau \longrightarrow \mu \gamma)$ vs $m_\Delta$.}
\label{fig:clfv}
\end{figure} 

\begin{figure*}
  \centering
  \subfigure[]{\includegraphics[width=0.33\textwidth]{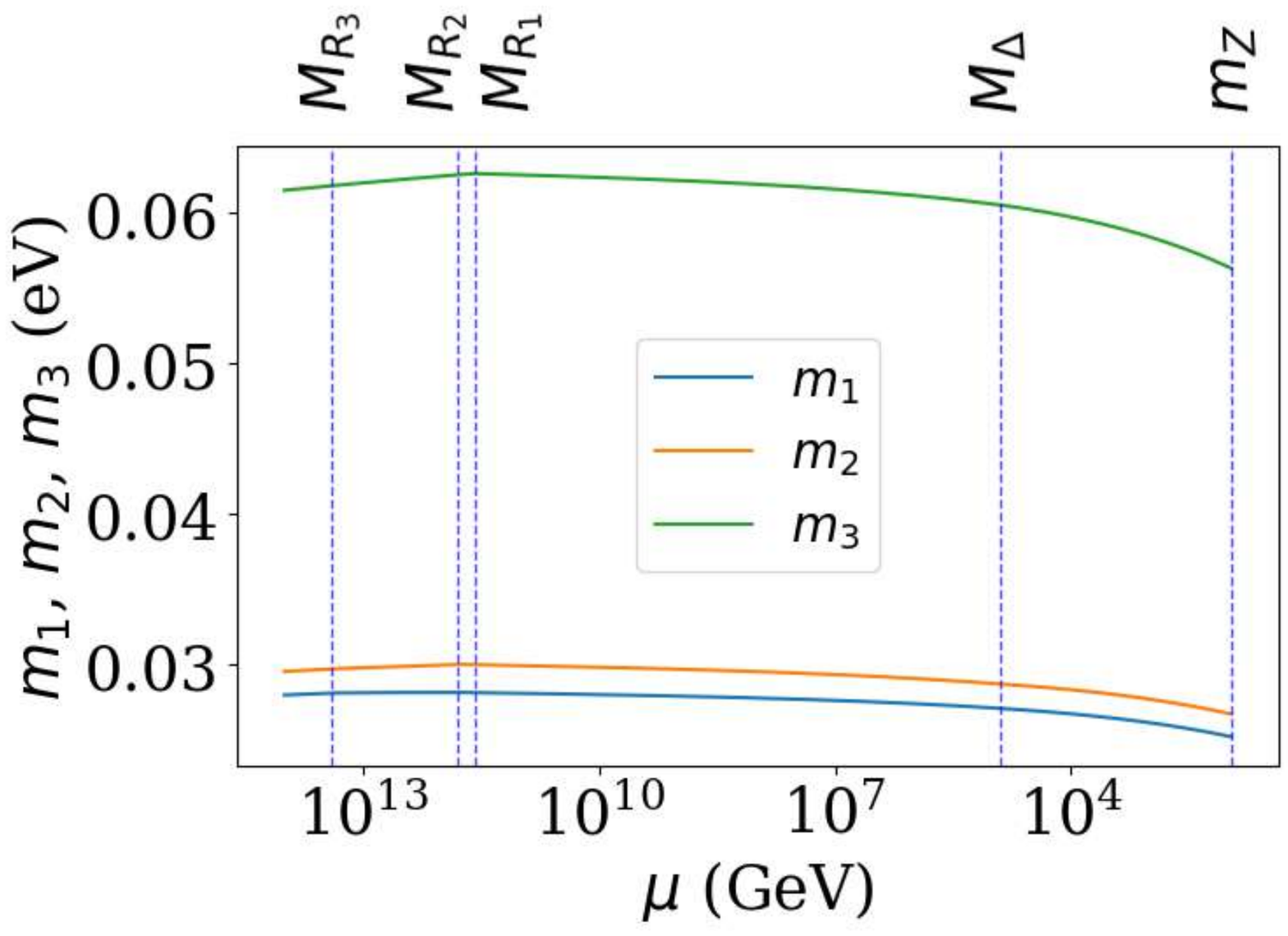}\label{fig:5(a)}}  
    \subfigure[]{\includegraphics[width=0.32\textwidth]{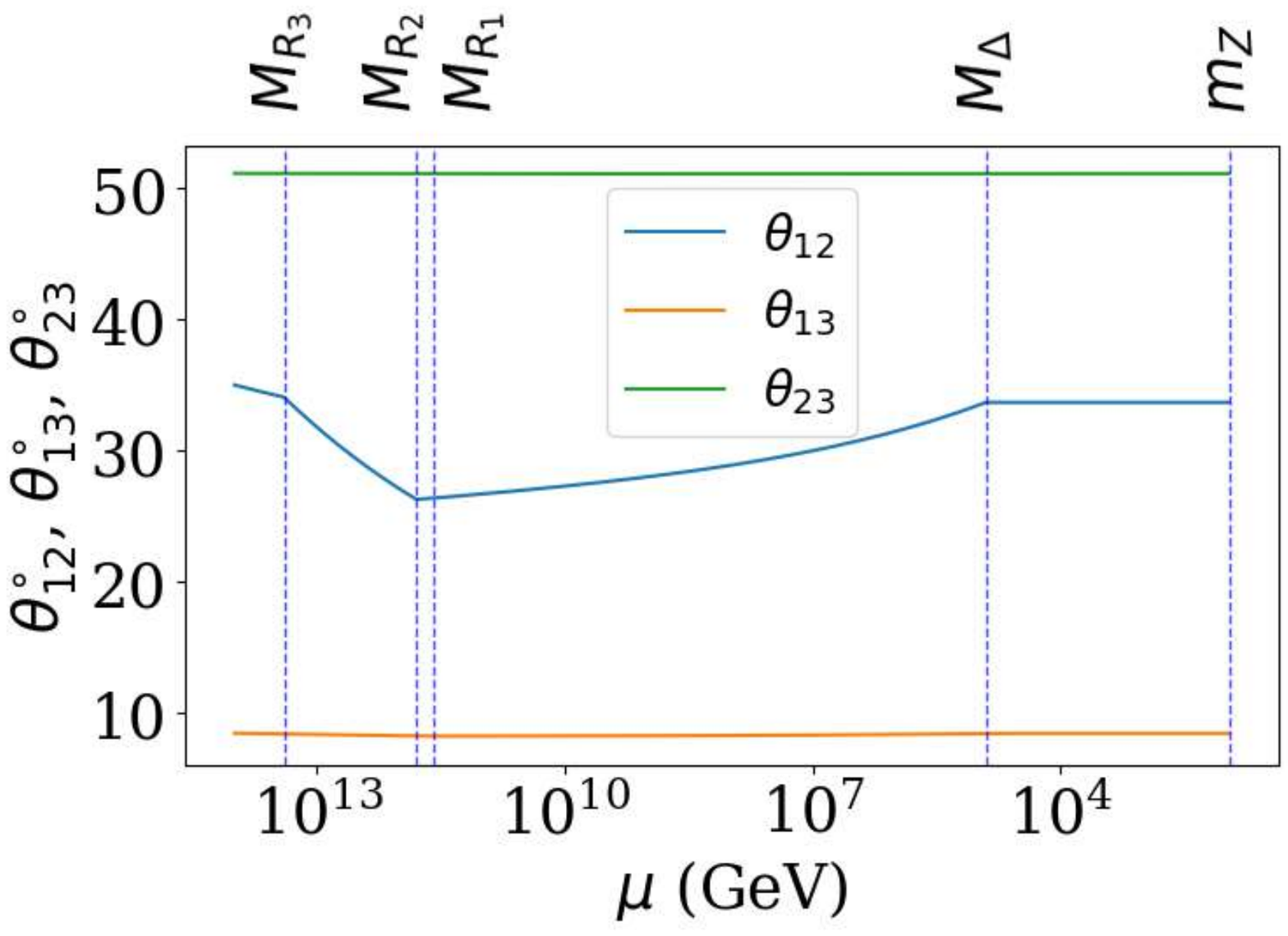}\label{fig:5(c)}}
    \subfigure[]{\includegraphics[width=0.33\textwidth]{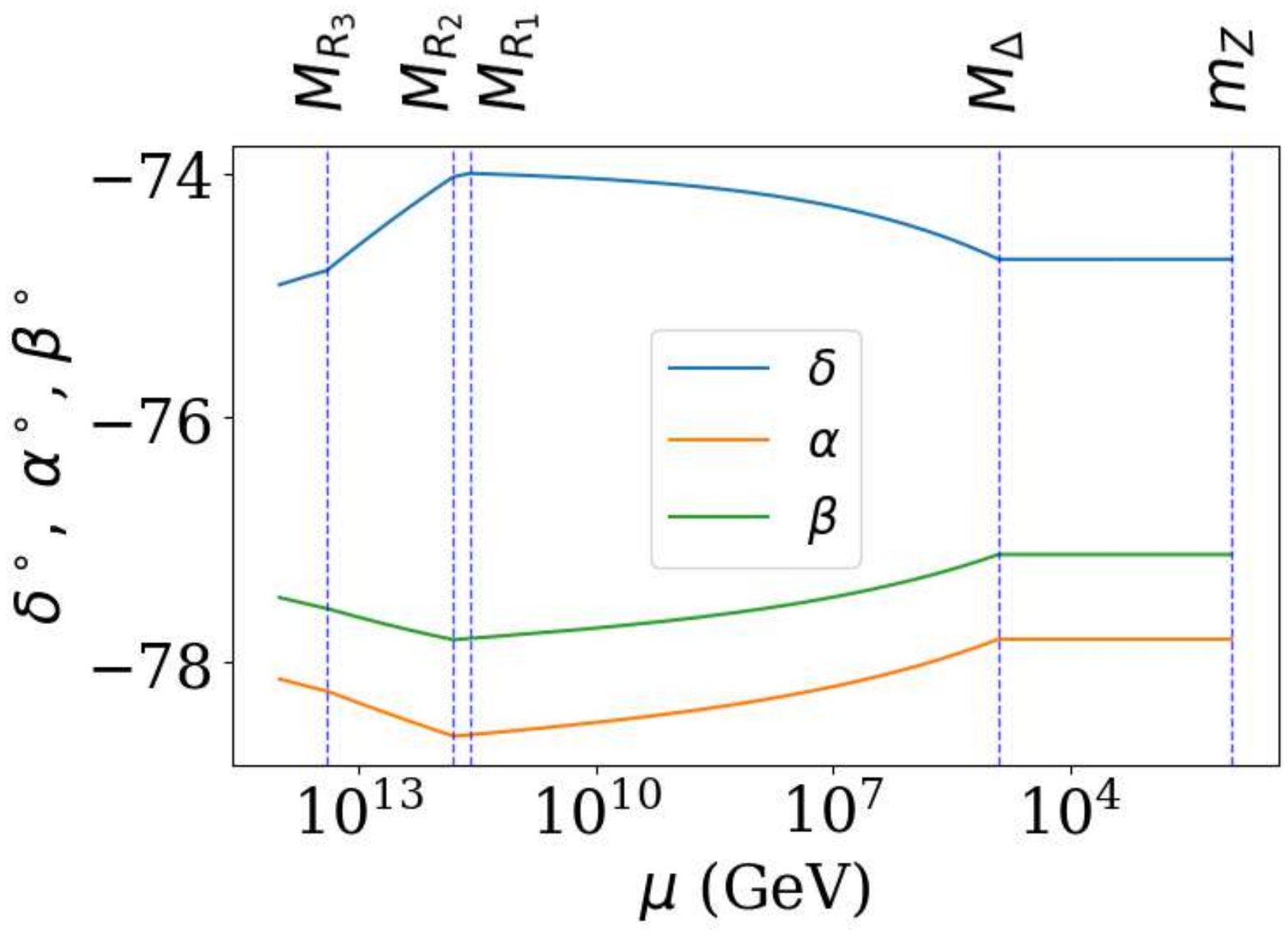}\label{fig:5(d)}}
    \caption{The RG evolution of the physical parameters from high energy scale $\Lambda$ to electroweak scale $m_Z$. The $M_{R_3}$, $M_{R_2}$, $M_{R_1}$ represent the masses of heavy right handed neutrinos and $m_\Delta$ stands for mass of scalar triplet.}
\label{fig:5}
\end{figure*}

In order to understand the possible origin of the posited texture, we consider a framework equipped with both type-I and type-II seesaw mechanisms\,\cite{Ma:2002nn}, which gives us the liberty to achieve prominent correlations among the mass matrix elements and ensures the independence of the four texture parameters while expressing them in terms of the model parameters. The Standard Model\,(SM) group is extended by $A_4 \times Z_{10} \times Z_{7} \times Z_{3}$ along with newly added scalar fields and right-handed neutrinos (see Table\,(\ref{Field Content of M})). The effective Lagrangian is shown below,   

\begin{eqnarray}
- \mathcal{L}_Y &=& \frac{y_{e}}{\Lambda^9} (\bar{D}_{l_{L}}H) e_{R} \xi^9 + \frac{ y_{\mu}}{\Lambda^6}(\bar{D}_{l_{L}}H)\mu_{R} \xi^6 + \frac{ y_{\tau}}{\Lambda^3}\nonumber\\&&(\bar{D}_{l_{L}}H)\tau_{R}\xi^3+ \,y_{1}(\bar{D}_{l_{L}}\tilde{H})\nu_{eR}+\frac{y_2}{\Lambda}(\bar{D}_{l_{L}}\tilde{H})\nonumber\\&&\nu_{\mu_R} \chi + \,\frac{y_3}{\Lambda} (\bar{D}_{l_{L}}\tilde{H})\nu_{\tau_R}\psi+ \frac{y_a}{2} (\bar{\nu}^c_{e_{R}}\nu_{e_{R}})\rho \nonumber\\&&+\frac{y_{b}}{2}[(\bar{\nu}^c_{\mu_{R}}\nu_{\tau_{R}}) + (\bar{\nu}^c_{\tau_{R}}\nu_{\mu_{R}})]\kappa+ \frac{y_{c}}{2}(\bar{\nu}^c_{\mu_{R}}\nonumber\\&&\nu_{\mu_{R}})\eta + \frac{y_{T}}{2} (\bar{D}_{l_{L}} D_{l_{L}}^{c})i \tau^2 \Delta + h.c.,
\label{Yukawa Lagrangian M1}
\end{eqnarray}

where, $\Lambda$ defines the cut-off scale of the theory. The presence of the auxiliary groups in the model restricts a few undesirable terms that are allowed by $A_4$. The group $Z_7$ forbids the appearance of Weinberg-like operators\,\cite{Vien:2025fiu, Chakraborty:2024hhq} in the framework. The group $Z_{10}$ primarily helps us to structure the right handed heavy neutrino mass matrix $M_R$. The group $Z_{3}$ is used to explain the charged lepton mass hierarchy. In the revised manuscript, we have included this part. Achieving the desired texture depends on an additional constraint, i.e., the proper alignment of the vacuum expectation values\,(vev) of the scalar field multiplets. We consider a specific choice: $\langle H \rangle_{0}=v_{H}(1,1,1)^{T}$, $\langle \Delta \rangle_{0}=v_{\Delta}(1,0,0)^{T}$, $\langle\chi\rangle_{0}=v_{\chi}$, $\langle\psi\rangle_{0}=v_{\psi}$, $\langle\xi\rangle_{0}=v_{\xi}$, $\langle\eta\rangle_{0}=v_{\eta}$, $\langle\kappa\rangle_{0}=v_{\kappa}$, and $\langle\rho\rangle_{0}=v_{\rho}$. The framework brings forth a non-diagonal charged lepton mass matrix, which is diagonalised with the help of $U_{l_L}$ and $U_{l_R}$, as shown below,
\begin{equation}
		U_{l_{L}}=\frac{1}{\sqrt{3}} \begin{bmatrix}
			e^{i\zeta} &  1 & 1\\
			e^{i\zeta} & \omega & \omega^2\\
			e^{i\zeta} & \omega^2 & \omega\\
		\end{bmatrix},\quad
		U_{l_{R}}=\text{diag}\begin{bmatrix}
			e^{i\zeta}, & 1, & 1
		\end{bmatrix}.
	\end{equation}

Here, we insist that the phase $\zeta$ appearing in $U_{l_{L}}$ and $U_{l_{R}}$ does not appear directly from the framework itself; rather, it is a consequence of the liberty to choose the eigenvectors in different possible ways\,\cite{Dey:2022qpu, Chakraborty:2023msb,Chakraborty:2024rgt, Chakraborty:2024hhq}. As expected, its presence does not alter the texture of $M_l$. However, in the light of charged lepton correction, the $U_{l_L}$ transforms the final version of $M_\nu$, $\zeta$ must therefore contribute to the latter. For our purpose, we fix $\zeta$ at $\pi/2$ and obtain the posited texture in Eq.\,(\ref{Mnu}) with,
\begin{eqnarray}
a&=&\frac{3 v_H^2 y_1^2}{y_a v_\rho}-\frac{2 v_\Delta y_T}{3},\quad g =\frac{6 v_H^2 y_3^2 y_c v_{\eta} v_{\psi}^2}{\Lambda^2 y_b^2 v_{\xi}^2}+\frac{2 v_\Delta y_T}{3},\nonumber \\
b&=& -\frac{1}{3} i v_\Delta y_T \quad h=-\frac{6 v^2 y_2 y_3 v_{\chi} v_{\psi}}{\Lambda^2 y_b v_{\xi}}-\frac{v_\Delta y_T}{3}.
\end{eqnarray}

The Yukawa couplings in the model are all assumed to be complex numbers. It is important to have an idea of the possible domain of the model parameters. For this model, we adhere to the naturalness of the Yukawa couplings, i.e., $|y_i| \sim \mathcal{O}(1)$ \,(see Fig.\,(\ref{fig:yukawa})). To achieve this realistic conditions, we stick to the values for the following parameters: $\Lambda \sim 10^{14}$ GeV, $v_\xi \sim 10^{18}$ GeV, $v_\psi \sim 10^{20}$ GeV, $v_\chi \sim 10^{19}$ GeV, $v_{\rho} \sim 10^{22}$ GeV, $v_\Delta \sim 0.10$ eV, $v_\eta \sim \times 10^{18}$ GeV and $y_{1,2,3}\sim \mathcal{O}(1)$. Needless to mention, $v_H$ is set at 246 GeV.

As an additional consequence, the framework addresses the hierarchy issue of the charged lepton masses through Froggat-Nielsen mechanism\,\cite{Froggatt:1978nt}: 	
\begin{eqnarray}
m_e : m_\mu : m_\tau &=& \Omega^6 y_e : \Omega^3 y_\mu : y_\tau,
\end{eqnarray}
where, $\Omega=v_\xi/\Lambda$.

Even though it is difficult at this stage to decipher the exact values of the model parameters, the model still gives us scope to estimate the mass of the scalar field $\Delta$ in the light of charged lepton flavour violation (CLFV) stringent constraints. The type-I seesaw contribution towards CLFV is suppressed because of the heavy right-handed neutrino mass\,\cite{Bilenky:1977du, Ilakovac:1994kj}. The model showcases that the branching ratios involving the decays $\mu \to e\gamma$, $\tau \to e \gamma$, or $\tau \to \mu\gamma$, which arise from scalar triplet exchange\,\cite{Ferreira:2019qpf}, are equal, and they are estimated as

\begin{align}
\mathrm{BR}(l \to l'\gamma)=
\frac{81 \alpha_{\rm em}}{192\,\pi\,G_F^2}\;
\;
\frac{\left| y_t \right|^4}{m_\Delta^4},
\end{align}

While deriving the above expression, we consider $\Delta$ scalar fields having equal masses $m_\Delta$. In the light of expected upper bound of the branching ratio of the ongoing MEG II experiment\,($<6\times 10^{-14}$, 90\% C.L.)\,\cite{Venturini:2024dvx}, and considering $y_t\sim \mathcal{O}(1)$, we obtain the mass of $\Delta$ as $m_\Delta \sim (5.69 \times 10^4 - 9.99 \times 10^4$)\,GeV\,(see Fig.\,\ref{fig:clfv}).

Since the model is realized at a high scale, it is essential to examine whether the resulting texture remains stable under renormalization group\,(RG) evolution down to the electroweak scale\,\cite{Schmidt:2007nq, Chankowski:1993tx, Grimus:2004yh}. In this regard, we perform a top-down one-loop RG analysis from $\Lambda$ to $m_Z$, consistently incorporating the sequential decoupling of heavy right-handed neutrinos as well as the scalar triplet threshold effects. We observe that the RG evolution of the physical parameters remain stable\,(see Fig.\,(\ref{fig:5})). The texture correlations remain approximately invariant under the RG evolution with $|M_{12}-M_{13}|\sim10^{-5}~\mathrm{eV}$ and $|M_{33}-2i M_{12}|\sim 10^{-4}~\mathrm{eV}$ respectively. 

In summary, we propose a predictive neutrino mass matrix with four complex parameters. The texture accounts for all observed neutrino parameters, rules out the inverted mass hierarchy, establishes the normal hierarchy by predicting the individual mass eigenvalues, and forbids the extreme cases where $m_1=0$. It precisely predicts $\theta_{23}$ and $\delta$ by placing them in the upper octant and fourth quadrant, respectively. In addition, the texture estimates the Majorana phases in the second quadrant. Further, we develop the neutrino mass matrix in exact form, maintaining the independence of the parameters, starting from a framework encompassing type-I and type-II seesaw, in the light of $SU(2)_L \times U(1)_Y \times A_4 \times Z_{10} \times Z_{7} \times Z_{3}$ symmetry. To see the potential consequences of the model, we further extend our discussion to $m_{\beta\beta}$, $A_{\mu e}$, and CLFV. In addition, we see that the proposed texture is approximately stable under the top-down RGE evolution. Although, the exact high-scale correlations receive mild corrections, the physical parameters remain stable under RG evolution. In this work, the charged lepton mass hierarchy is explained through the Froggatt-Nielsen mechanism\,\cite{Froggatt:1978nt}. At first glance, it may seem misleading that the Yukawa terms of the neutrino sector are suppressed by $\sim 1/\Lambda, 1/\Lambda^2$, whereas for charged leptons, the suppressions are of the order $1/\Lambda^{9}$. Here, it is important to note that the involvement of several vevs plays a crucial role in addressing this anomaly and thus naturally explaining the desired results\,\cite{Chakraborty:2025shb}. We highlight that the presence of auxiliary groups restricts other possibilities of $1/\Lambda^n$, with $n > 9$, for the charged lepton sector from appearing in the theory. 

\section*{Acknowledgements}
The research work of PC is supported by Innovation in Science Pursuit for Inspired Research (INSPIRE), Department of Science and Technology, Government of India, New Delhi vide grant No. IF190651.

\biboptions{sort&compress}
\bibliography{ref.bib}

\end{document}